\documentclass[aip,reprint,amsfonts,amsmath,amssymb,floatfix,citeautoscript,groupedaddress]{revtex4-1}
\usepackage{bm}
\usepackage{dcolumn}
\usepackage{graphicx}
\usepackage{parskip}
\usepackage{color}
\usepackage{enumerate}
\usepackage{psfrag}
\usepackage{upgreek}
\setcitestyle{notesep={ }}
\setcitestyle{numbers,square}

\newcommand{\citesx}{\setcitestyle{open={[},close={,}}\cite}
\newcommand{\citedx}{\setcitestyle{open={},close={]}}\cite}
\newcommand{\citevv}{\setcitestyle{open={[},close={]}}\cite}

\begin{document}

\title{
On the $in$existence of a ``splay(-bend)'' nematic phase
}


\author{Giorgio Cinacchi}
\affiliation{
Departamento de F\'isica Te\'orica de la Materia Condensada,  \\
Instituto de F\'isica  de la Materia Condensada (IFIMAC),    \\
Instituto de Ciencias de Materiales ``Nicol\'{a}s Cabrera'' (INC), \\
Universidad Aut\'onoma de Madrid, \\
Ciudad Universitaria de Cantoblanco, 
E-28049 Madrid, Spain
}


\begin{abstract}
On the basis of 
the application of the (Onsager) second-virial density functional theory to 
an artificial system that is so designed as to be the best promoter of a ``splay(-bend)'' nematic phase,
it is argued that this ``modulated'' nematic phase cannot exist.
\end{abstract}

\maketitle

In these years, there has been much ado about ``modulated'' nematic liquid-crystalline phases.
These phases are, or would be, liquid-crystalline in the sense that they are \textit{fluid}.
These phases are, or would be, nematic 
in the sense that 
their constituent (molecular, colloidal, granular) particles 
\textit{do not} have \textit{long}-distance positional order 
but 
\textit{do} have \textit{long}-distance orientational order:
the (primary) axes of the particles preferentially align along 
a directional axis that is traditionally 
denominated the (primary) director.
These phases are, or would be, ``modulated'' nematic 
in the sense that 
the (primary, secondary) director, 
instead of being fixed in space as in the non-``modulated'', ordinary, nematic phase,
regularly translo-rotates along and round another directional axis:
the type of translo-rotation differentiates among the ``modulated'' nematic phases.

\begin{figure}[h!]
\centering
\includegraphics[scale=0.775]{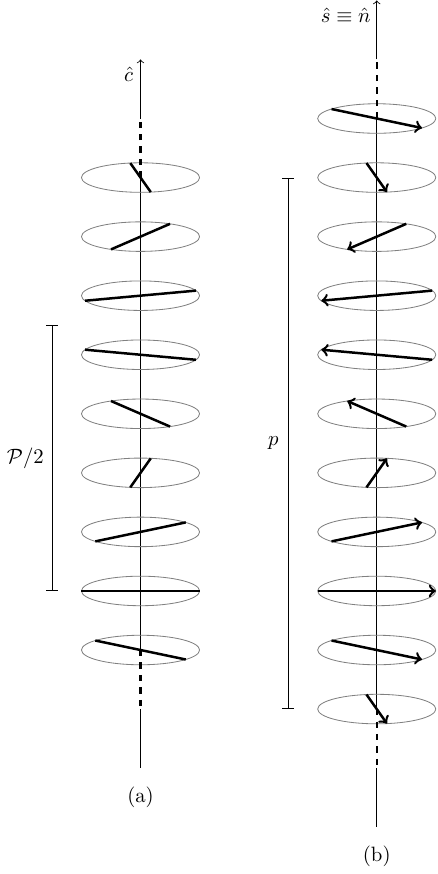}
\caption{Schematic of the variation of 
(a) \(\hat n\) along and round the cholesteric axis \(\hat c\) 
in a cholesteric phase with algebraic pitch \({\cal P}/2\), Eq. \ref{equ1},
and
(b) \(\hat{\mathbf m}\) along and round the screw axis \(\hat s\) 
which coincides with the nematic director \(\hat n\) 
in a screw nematic phase with algebraic pitch \(p\), Eq. \ref{equ2}.}
\label{figurameno1}
\end{figure}

One ``modulated'' nematic liquid-crystalline phase is common and long known:
the cholesteric phase \cite{straley,priestley,blinov}. 
It can form in a three--dimensional system of 
not only sufficiently non-spherical but also chiral
particles. 
It is such that these particles preferentially align their primary axes 
along the primary nematic director \(\hat n\)
which dextro(sinistro) translo-rotates along and round 
the cholesteric axis \(\hat{c}\).
If, e.g., \(\hat{c}\, \| \, {\mathbf{\hat {z}}}\) then:
\begin{equation}
{\hat n} (x\,,y\,,z) = 
\begin{cases}
n_{x} = \pm \sin \left( 2\, \uppi \,z/{\cal P}\right) \\[3pt]
n_{y} = \pm \cos \left( 2\, \uppi \,z/{\cal P}\right) \\[3pt]
n_{z} = 0
\end{cases}
\label{equ1}
\end{equation}
with the use of \(\hat n\) and the consequent use of the \(\pm\) sign 
that signify the local nematic symmetry of the phase and 
with \({\cal P}/2\) the algebraic cholesteric phase pitch
[Fig. \ref{figurameno1}(a)].

One more ``modulated'' nematic liquid-crystalline phase is more recent and special: 
the screw nematic phase \cite{nemavite1,nemavite2,nemavite3}.
It can form in a three--dimensional system of sufficiently elongate helical particles with
helical axis \(\hat h\), algebraic pitch \(p\) and radius \(r\).
It is such that these particles preferentially align their helical axes along 
the primary nematic director \(\hat n\)
which is fixed in space as in the non-``modulated'', ordinary, nematic phase 
but 
there is a secondary polar director \(\hat{\mathbf m}\) 
which dextro(sinistro) translo-rotates along and round the screw axis \({\hat s} \equiv {\hat n}\).
If, e.g., \({\hat s} \equiv {\hat n} \, \| \, {\hat{\mathbf z}} \) then:
\begin{equation}
{\hat {\mathbf m}} (x\,,y\,,z) = 
\begin{cases}
m_{x} = + \sin \left( 2\, \uppi \,z/{p}\right) \\[3pt]
m_{y} = + \cos \left( 2\, \uppi \,z/{p}\right) \\[3pt]
m_{z} = 0
\end{cases}
\label{equ2}
\end{equation}
with the use of \(\hat{\mathbf m}\) and the consequent use of the \(+\) sign that
signify the local polar symmetry of the phase and 
with \(p\) also the algebraic screw nematic phase pitch
[Fig. \ref{figurameno1}(b)].

The last fifty years have seen how 
the conjecture that other ``modulated'' nematic liquid-crystalline phases could form 
in systems of non-chiral but polar particles \cite{meyer,pleiner}, 
such as systems of arcuate or bent particles or of cuneate particles,
has been progressively consolidating into, 
first, explicit theoretical predictions of their existence and, 
then,  longed-for experimental claims of their observation. 

\begin{figure}
\centering
\includegraphics[scale=0.875]{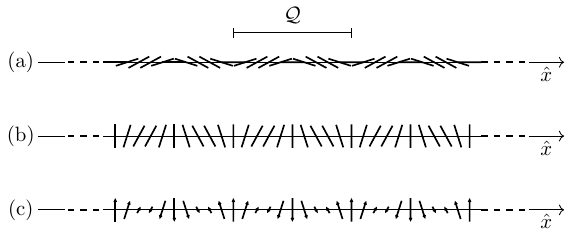}
\caption{Schematic of the variation of (a) \(\hat n\) along the directional axis \(\hat x\)
in the would-be ``splay-bend'' nematic phase, Eq. \ref{equ3}, 
(b) \(\hat n\) along the directional axis \(\hat x\)
in the would-be ``splay'' nematic phase, Eq. \ref{equ4}, 
and (c) \(\mathbf p\) along the directional axis \(\hat x\) in both
the would-be ``splay(-bend)'' nematic phases, Eq. \ref{equ5},
being in (a,b,c)  \(\vartheta \stackrel{{\rm e.g.}}{=} \uppi/6\) and
\(\cal Q\) the ``wave''length of the ``splay(-bend)'' periodicity. }
\label{figura0}
\end{figure}

One of these ``modulated'' nematic liquid-crystalline phases has been ultimately denominated
``splay(-bend)'' nematic phase \cite{dozov,goodby}. 
It would form in a system of polar particles, 
such as a system of arcuate or bent particles or of cuneate particles.
It would be such that these particles preferentially align 
certain of their axes (the tangential axes; the two-fold symmetry axis) along a primary nematic director \(\hat n\) 
which, longitudinally or laterally, serpentines along a directional axis in a plane.
If, e.g., the directional axis is \(\hat{x}\,\|\,\hat{\mathbf{x}}\) and 
the plane is the (\textsl{x},\textsl{y}) plane 
then it was assumed that \cite{dozov}:
\begin{equation}
{\hat n} (x\,,y\,,z) = 
\begin{cases}
n_{x} = \pm \cos \left( \vartheta \sin (2\, \uppi \,x/{\cal Q}\right) ) \\[3pt]
n_{y} = \pm \sin \left( \vartheta \sin (2\, \uppi \,x/{\cal Q}\right) ) \\[3pt]
n_{z} = 0
\end{cases}
\label{equ3}
\end{equation}
[Fig. \ref{figura0}(a)] or it has been also assumed that \cite{goodby}:
\begin{equation}
{\hat n} (x\,,y\,,z) = 
\begin{cases}
n_{x} = \pm \sin \left( \vartheta \sin (2\, \uppi \,x/{\cal Q}\right) ) \\[3pt]
n_{y} = \pm \cos \left( \vartheta \sin (2\, \uppi \,x/{\cal Q}\right) ) \\[3pt]
n_{z} = 0
\end{cases}
\label{equ4}
\end{equation}
[Fig. \ref{figura0}(b)] with the use of \(\hat n\) that would oblige the use of the \(\pm\) sign 
in accordance to the local nematic symmetry of the phase and with 
\(\vartheta\) a certain characteristic angle that contributes to setting the extent
of the ``splay(-bend)'' ``modulation'' and 
\(\cal Q\) either 
the ``wave''length of the ``splay-bend'' periodicity 
in Eq. \ref{equ3} \cite{dozov} [Fig. \ref{figura0}(a)]  
or 
the ``wave''length of the ``splay'' periodicity 
in Eq. \ref{equ4} \cite{goodby} [Fig. \ref{figura0}(b)].

\smallskip
Naturally, as much as calamitic nematics are equivalent to discotic nematics in three dimensions or,
mutatis mutandis, 
as much as there cannot be any distinction between calamitic nematics and discotic nematics in two dimensions,
a ``splay-bend'' nematic would be equivalent to a ``splay'' nematic \cite{kamien}.  

Since the particles that would form such a ``splay(-bend)'' nematic phase are arcuate or bent or they are cuneate,
they are polar.
There exists a polar director \(\hat{\mathbf p}\), 
which actually \textit{is} the primary director of such a conjectural ``modulated'' nematic phase,
along which the arcuate or bent particles or the cuneate particles preferentially align their polar axes.
Irrespective of the form that has been assumed for \(\hat n\), 
Eq. \ref{equ3} or Eq. \ref{equ4},
the polar vector \({\mathbf p} = |\mathbf{p}|\,\hat{\mathbf{p}}\) 
laterally oscillatorily varies along \(\hat{x}\)
\cite{dozov} 
and has been assumed as \cite{goodby,kamien}:
\begin{equation}
{{\mathbf p}} (x\,,y\,,z) = 
\begin{cases}
p_{x} = + P \cos(2\, \uppi \,x/{\cal Q}) \sin \left( \vartheta \sin (2\, \uppi \,x/{\cal Q}\right) ) \\[3pt]
p_{y} = + P \cos(2\, \uppi \,x/{\cal Q}) \cos \left( \vartheta \sin (2\, \uppi \,x/{\cal Q}\right) ) \\[3pt]
p_{z} = 0
\end{cases}
\label{equ5}
\end{equation}
with the use of \({\mathbf p}\) and the consequent use of the \(+\) sign that
signify the local polar symmetry of the phase and with
\(P\) a constant that, if  \(0 \leq P \leq 1\), 
makes \( \mathtt{S}_1(x)=P \cos(2\, \uppi \,x/{\cal Q}) \) become the local polar order parameter,
which is equal to \((-1)^j\,P\) at \(x=j{\cal Q}/2\), \(j\in\mathbb{Z}\), 
and equal to \(0\) at \(x=(2j+1)\,{\cal Q}/4\), \(j\in\mathbb{Z}\) 
[Fig. \ref{figura0}(c)].

Naturally, to qualify for being nematic a phase must be positionally disordered.
Here, that a phase can be positionally disordered \textit{and} could have a 
suitably laterally oscillatory
polar director
is questioned and 
physical counterarguments are provided 
that indicate that a ``splay(-bend)'' nematic phase cannot exist:
either the phase is a nematic phase but not ``splay(-bend)'' 
or the phase is a cluster phase
but not nematic.

To provide physical counterarguments that are as firm as possible,
one 
sets in a situation that arguably is the most favourable 
to the formation of such a conjectural ``modulated'' nematic phase and
values whether it can ever form in this most favourable situation.

First, 
one sets in two dimensions
without any loss of generality. 
Even though the ``splay(-bend)'' nematic phase was 
predicted and claimed to have been observed 
in three dimensions \cite{dozov,goodby},
a cursory view of Eqs.  \ref{equ3}, \ref{equ4}, \ref{equ5} suffices to recognize 
that it effectively is a phase in two dimensions:
in the right-hand side of Eqs. \ref{equ3}, \ref{equ4}, \ref{equ5} the variable \(z\) is absent and 
\(n_z=0\) in Eqs. \ref{equ3} and \ref{equ4} and \(p_z = 0\) in Eq. \ref{equ5}.

Second, one artificially deliberately removes
the isotropic phase and the non-``modulated'', ordinary, nematic phase.
Naturally, any liquid-crystalline phase has to compete for its existence against the ever-present isotropic phase: 
it is always there to erode the region of stability of any liquid-crystalline phase from below in density.
Naturally, a non-``modulated'', ordinary,  nematic phase can form in systems of sufficiently non-spherical polar particles:  
it can be there to erode the region of stability of a ``modulated'' nematic phase from below in density.
To artificially yet most severely contrast this  action of erosion
of the isotropic phase and of the non-``modulated'', ordinary, nematic phase, 
it is artificially assumed perfect local polar (nematic) order along a
suitably laterally oscillatory
polar (nematic) director:
the arcuate or bent particles or the cuneate particles perfectly align their polar axes 
along the local polar director \(\hat{\mathbf{p}}\) 
that suitably laterally oscillates
along \(\hat{x}\).
In fact, it is possible to imagine 
a perfect non-``modulated'', ordinary, nematic phase 
as well as a perfect cholesteric phase 
as well as a perfect screw nematic phase
in which the particles perfectly align along the respective (primary, secondary; polar, nematic) directors 
and these perfect orientational orders are compatible with 
positional disorder at sufficiently low but finite density.
Here, it is set perfect local polar (nematic) order along a 
suitably laterally oscillatory
polar director and
it is inquired whether this perfect orientational order is compatible with positional disorder
at sufficiently low but finite density.

\begin{figure}
\centering
\includegraphics[scale=1.00]{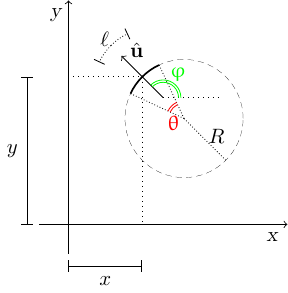}
\caption{
Example of a minor circular arc (continuous) in the {\rm (\textsl{x},\textsl{y})} plane.
The minor circular arc is with a subtended angle \(\uptheta \in [0,\uppi]\) (red or dark gray) and 
has been severed from a circumference of radius \(R\) (discontinuous) so that
its length \(\ell = \uptheta\,R\) (discontinuous).
Its vertex has position (\(x\),\(y\)) and 
its orientation is defined by the unit vector \(\hat{\mathbf{u}}\) 
which is in the direction and sense 
from the center of the circumference to the vertex of the minor circular arc and 
forms an angle \(\upvarphi\) (green or light gray) with the {\rm \textsl{x}} axis. 
}
\label{figura1}
\end{figure}

Third, one considers what arguably is the most appropriate elementary arcuate or bent particle model:
hard, infinitesimally--thin, minor circular arcs (Fig. \ref{figura1}).
These curved particles are identified by the subtended angle \(\uptheta \in [0,\uppi] \) and 
have been severed from a circumference of radius \(R\) 
so that their length \(\ell = \uptheta \,R\) 
(Fig. \ref{figura1}).
They are naturally infinitesimally--thin (Fig. \ref{figura1}).
These particles are hard in that 
they cannot intersect but else do not interact.
Here, it is shared the conviction  that 
``excluded-volume'' interactions, i.e., packing effects, are 
the most basic effects and interactions in classical (dense) systems of particles \cite{torquato}; 
a conviction that, 
if it were indeed shared, 
should compel the use of hard-particle models,
hence discrete-particle models,
in the attempts to understand the statistical physics of these systems.
The fact that they are hard and infinitesimally--thin and minor circular makes 
these particles the best promoters of (local) nematicity against 
the formation of other (liquid-)crystalline phases,
as much as 
hard infinitesimally--thin discs in three dimensions \cite{harddiscs} and
hard linear segments in two dimensions \cite{hardsegments}  
can only form a (quasi-)nematic phase and 
no positionally (partially) ordered phase 
as they are progressively compressed at finite density.

Therefore, the artificial system that one has to consider is
a system of hard minor circular arcs 
which move in such a manner that 
their polar axes always perfectly point along
the suitably translo-rotating local polar director 
in the two--dimensional Euclidean space.

\begin{figure}
\centering
\includegraphics[scale=1.10]{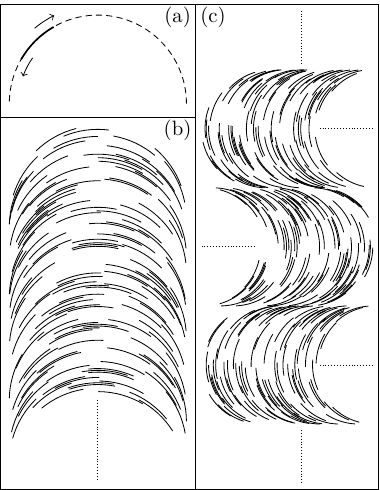}
\caption{(a) The translo-rotational motion of a hard minor circular arc on the parent semicircumference.
(b) In addition to this translo-rotational motion, hard minor circular arcs move along the polar axis of 
their parent semicircumferences thus generating a filament.
(c) Filaments succeed side-left by side-right thus generating the artificial system.}
\label{figura2}
\end{figure}

\begin{figure}
\centering
\includegraphics[scale=0.9]{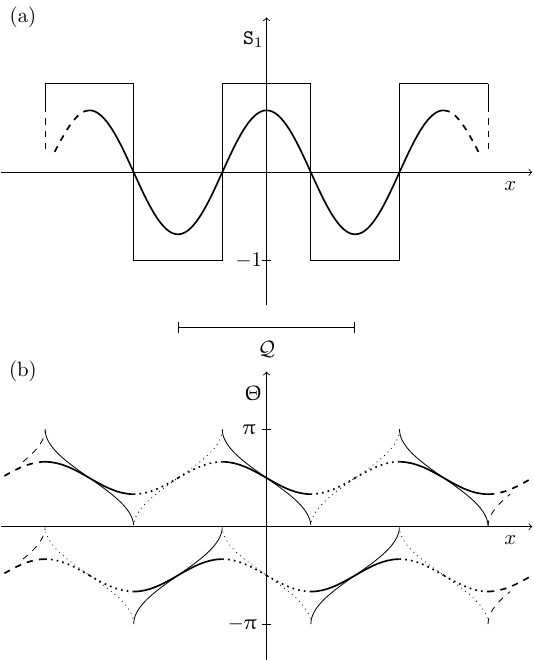}
\caption{The behaviour of (a)
\(\mathtt{S}_1\), the local polar order parameter, 
and  
(b) \( \Uptheta = \arccos({\hat{\mathbf{p}}}\, \cdot \, {\hat {\mathbf {x}}}) \), 
the angle that the polar vector \(\mathbf{p}\) makes with the directional axis, 
as a function of \(x\), 
the position along the directional axis, 
from Eq. \ref{equ5} (thick solid line), taking  \(P\stackrel{\rm e.g.}{=}0.7\) in (a) and 
\(\vartheta\stackrel{\rm e.g.}{=}\uppi/6\) in (b), 
and from Eq. \ref{equ6} (thin solid line),
once a many-filament system is considered and then the domains of
\(\mathtt{S}_1(x)\) and \(\Uptheta(x)\) have been suitably extended and polarity alternation, as in (a), has been introduced;
because of polarity alternation in (a) the solid lines
in (b) alternate between the half-plane with positive
values of \(\Uptheta\) and the half-plane with negative
values of \(\Uptheta\);
the dashed tracts in (a,b) and the dotted tracts in (b) would want to signify curve continuation.}
\label{figura3}
\end{figure}

On the basis of previous results on both 
systems of hard spherical caps \cite{hardsphericalcaps}, 
the three--dimensional analogue of systems of hard minor circular arcs,
and 
systems of hard minor circular arcs \cite{hardarcs},
one can realize that such a particular motion that
a hard minor circular arc realizes consists in
that translo-rotational motion that 
allows it to perfectly adapt to a parent semicircumference:
a sort of glide and rock on a dome [Fig. \ref{figura2}(a)].
Therefore, the artificial system of hard minor circular arcs
consists of such hard particles translo-rotating along parent semicircumferences
as well as translating along the polar axes of these parent semicircumferences,
thus resulting to form sort of filaments [Fig. \ref{figura2}(b)].
The polar axes of the parent semicircumferences 
in the filaments can alternately point 
so that the system is globally non-polar and
thus, for this artificial system,
the ``wave''length of the periodicity is \({\cal Q}=4R\) [Fig. \ref{figura2}(c)].
Here, one has to note that this alternation is actually an ineffective assumption.
In fact, in this artificial system, the filaments are independent from one another
as hard minor circular arcs that belong to two adjacent filaments cannot intersect. 
Therefore, one can focus on one representative filament.
If the axis of the representative filament 
is arbitrarily taken along the \textsl{y} axis,
that the hard minor circular arcs can perfectly point their polar axes along the local polar director
implies that the local polar director and the local polar vector,
in the representative filament, actually is:
\begin{equation}
{\mathbf p} ({x}) =
{\hat{\mathbf p}} ({x}) =
\begin{cases}
p_{x} = \dfrac{{x}}{R}  \\[5pt]
p_{y} = \sqrt{1-p_{x}^2}  
\end{cases}
-R \leq x \leq R  \,.
\label{equ6}
\end{equation}
It may be pertinent to note that 
the polar director profile in Eq. \ref{equ5} 
clearly differs from 
the polar director profile in Eq. \ref{equ6} 
in several respects.
Eq. \ref{equ5} amounts to taking an imperfect local polar order,
particularly a value of \(\mathtt{S}_1\) such that
\(\max\,\left|\mathtt{S}_1 (x) \right| = P \leq 1\), and 
that varies along the directional axis as a sine wave,
while Eq. \ref{equ6} is consistent with a perfect local polar order,
i.e., a value of \(\mathtt{S}_1\) such that
\(\left|\mathtt{S}_1 (x) \right| = 1\), and 
that varies along the directional axis as a square wave
[Fig. \ref{figura3}(a)]. 
Eq. \ref{equ5} assumes 
that the angle \(\Uptheta\) 
which the polar (nematic) director forms with the directional axis 
varies sinusoidally, 
while Eq. \ref{equ6} implies 
that the angle \(\Uptheta\)
which the polar (nematic) director forms with the directional axis
varies arc-cosinusoidally
[Fig. \ref{figura3}(b)]. 
Both Eq. \ref{equ5} and  Eq. \ref{equ6} are ans\"atze:
there is no basic reason to prefer either except that
Eq. \ref{equ5} seems to appeal to generic, algebraic conveniences, 
while Eq. \ref{equ6} results from specific, geometric, considerations.

One has to note that all these considerations are valid irrespective of the system being 
uni-component, i.e., composed of 
hard minor circular arcs with the same \(\uptheta\) and \(R\) and hence \(\ell\) or
multi-component, i.e., composed of
hard minor circular arcs with the same \(R\) but different \(\uptheta\) and hence
different \(\ell\).

Therefore, one has to investigate such an artificial, effectively one-filament, generally multi-component, system.

Each component hard minor circular arc of type \(k\) 
is identified by its subtended angle \(\uptheta_k\)
with its length \(\ell_k=\uptheta_k\,R\).
For simplicity, but essentially without any loss of generality,
such an artificial system can be considered as composed of 
a discrete number n of component hard minor circular arcs. 
Naturally, the composition of such an artificial system can be defined by the set of number fractions 
\(\displaystyle \left\{\mathtt{x}_k\right\}_{\rm n} = \left\{\mathtt{x}_k\right\},\,k=1,\dots{\rm n}\), 
which are such that \(\displaystyle \sum_{k=1}^{\rm n} \mathtt{x}_k = 1\);
the set \(\mathtt{X}\) of all the compositions can be defined as 
\(\displaystyle \mathtt{X} = \left\{\forall {\rm n}\in \mathbb{N}\,, 
\mathtt{x}_1,\dots,\mathtt{x}_{\rm n}\in\mathbb{R}^{+}\,: 
\sum_{k=1}^{\rm n} \mathtt{x}_k = 1\right\}\).
Therefore, one has to investigate such an artificial system 
\(\forall \left\{\mathtt{x}_k\right\}_{\rm n} \in \mathtt{X}\)
as a function of the number density \(\uprho\) and, \(\forall k,\, k=1,\dots,{\rm n}\),
determine the positional distribution functions \(G_k(x)\),
which can be normalized as
\begin{equation}
\dfrac{1}{2R} \int_{-R}^{+R} dx\,G_k(x) = 1 \,,
\label{equnorma}
\end{equation}
so that \(dx\,G_k(x)\,/(2R)\) provides 
the probability to find a hard minor circular arc of type \(k\)
whose vertex has position in the interval \([x, x+dx]\).

The basic question to pose now becomes:
\(\forall \left\{\mathtt{x}_k\right\}_{\rm n} \in \mathtt{X}\), 
what is the form of \(G_k(x)\), \(\forall k,\, k=1,\dots,{\rm n}\), as a function of \(\uprho\)?
If \( \exists \left\{\mathtt{x}_k\right\}_{\rm n} \in \mathtt{X} \) for which
\(G_k(x) = 1\), \(\forall k, \, k=1,\dots,{\rm n}\), at a certain value
of \(\uprho\) then this would correspond to the ``splay(-bend)'' nematic phase.
Can the artificial system of hard minor circular arcs, 
that has been so designed as to be the best promoter of the ``splay(-bend)'' nematic phase, 
be such that \( \exists \left\{\mathtt{x}_k\right\}_{\rm n} \in \mathtt{X} \) 
for which  \(G_k(x)=1\), \(\forall k, \, k=1,\dots,{\rm n}\), in a finite interval of \(\uprho\)?

For any composition \(\left\{\mathtt{x}_k\right\}_{\rm n} \in \mathtt{X}\),
the determination of \(G_k(x)\), \(\forall k, \, k=1,\dots,{\rm n}\), as a function of \(\uprho\) 
admittedly requires a discrete particle-based method.
In particular, one thinks of numerical simulation,
the only truly predictive and reliable theoretical method 
that one has at disposal \cite{allen}.
It is feasible to set up 
a Monte Carlo numerical simulation \cite{allen}
of the artificial system under consideration.
However, in the physics of liquid crystals, 
there is a more expeditious method 
which is still discrete particle-based and,
if it is used in the appropriate situations, 
can provide results of a high degree of predictivity and reliability:
(Onsager) second-virial density functional theory \cite{onsager}.
Together with three--dimensional positionally disordered systems of hard elongate particles 
in the limit of infinite length and/or infinitesimal width \cite{onsager}, 
perfectly orientationally ordered systems of hard infinitesimally--thin particles are 
the best situations in terms of predictivity and reliability 
in which a second-virial density functional theory can be applied.

\begin{figure}
\centering
\includegraphics[scale=1.3]{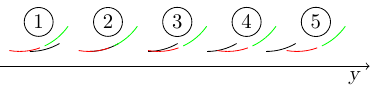}
\caption{
The geometric interpretation of the function \({\cal S}_{k_1\,k_2}(x|{\hat{\mathbf p}}(x)\,;\,x'|{\hat{\mathbf p}}(x'))\)
and its, equal to or larger than, 0  value.
From left to right, 
consider the five images that are each formed 
by a black minor circular arc and
by a green (light gray) and a red (dark gray) minor circular arc,
all minor circular arcs having, e.g., the same subtended angle \(\uptheta\), 
while the green (light gray) and the red (dark gray) minor circular arcs belong to the same parent semicircumference. 
The black minor circular arc and the green (light gray) minor circular arc are such that, respectively, 
\(|x| < R\cos(\uptheta/2)\) and  \(|x'| < R\cos(\uptheta/2)\) while
the red (dark gray) minor circular arc is such that \(|x'| > R\cos(\uptheta/2)\).
Imagine to 
maintain the black minor circular arc fixed and 
move the green (light gray) and red (dark gray) minor circular arcs 
along the {\rm \textsl{y}} axis and 
focus on the intersection between the black minor circular arc and 
either of the other two minor circular arcs.
While the black minor circular arc and the green (light gray) minor circular arc intersect
only when the parent semicircumferences coincide (image {\large{\textcircled{\small 2}}}),
the black minor circular arc and the red (dark gray)  minor circular arc can intersect
also if the two parent semicircumferences do not coincide
(images {\large{\textcircled{\small 2}}}, {\large{\textcircled{\small 3}}}, {\large{\textcircled{\small 4}}}):
for the pair black minor circular arc --- green (light gray) minor circular arc 
\({\cal S}(x|{\hat{\mathbf p}}(x)\,;\,x'|{\hat{\mathbf p}}(x'))=0\)
while for the pair black minor circular arc --- red (dark gray) minor circular arc
\({\cal S}(x|{\hat{\mathbf p}}(x)\,;\,x'|{\hat{\mathbf p}}(x'))>0\).
}
\label{figura4}
\end{figure}

For the artificial system of N hard minor circular arcs under consideration, 
with a certain composition  \(\left\{\mathtt{x}_k\right\}_{\rm n} \in \mathtt{X}\) and 
with a certain value of \(\uprho\), 
the free energy \(\cal F\) per particle, \(\mathsf{f} = {\cal F}/{\rm N}\), 
in the second-virial density functional theory approximation, 
is:
\begin{widetext}
\begin{eqnarray}
\upbeta {\mathsf{f}} \,
=\,&\log{\cal V}& + \log \uprho - 1 + 
\sum_{k=1}^{\rm n} \mathtt{x}_k \,\log \mathtt{x}_k +
\sum_{k=1}^{\rm n} \mathtt{x}_k \dfrac{1}{2R} \int_{-R}^{+R} dx\, G_k(x|{\hat{\mathbf p}}(x)) \log G_k(x|{\hat{\mathbf p}}(x)) + \nonumber \\
&+&\dfrac{1}{2} \uprho \sum_{k_1=1}^{\rm n} \sum_{k_2=1}^{\rm n} \mathtt{x}_{k_1} \mathtt{x}_{k_2} 
\dfrac{1}{2R} \int_{-R}^{+R} dx\, G_{k_1}(x|{\hat{\mathbf p}}(x)) 
\int_{-R}^{+R} dx'\, G_{k_2}(x'|{\hat{\mathbf p}}(x'))
{\cal S}_{k_1\,k_2} (x|{\hat{\mathbf p}}(x)\,;\,x'|{\hat{\mathbf p}}(x')) \,.
\label{equf}
\end{eqnarray}
\end{widetext}
\noindent In this Eq. \ref{equf}: 
\(\upbeta = 1/(k_B T)\), with \(k_B\) the Boltzmann constant and \(T\) the absolute temperature, and
\( \cal V\) the ``thermal'' area;
it has been specified that the function \(G_k\) depends 
on the variable \(x\) both explicitly and implicitly
via \(\hat{\mathbf p}(x)\); 
the function \({\cal S}_{k_1\,k_2}\),
which also depends on the variables \(x\) and \(x'\) 
both explicitly and implicitly via \(\hat{\mathbf p}(x)\) and \(\hat{\mathbf p}(x')\), is key and, 
like the analogue functions for other systems of hard particles, has a specific geometric interpretation:
it is the length of that segment parallel to the  \textsl{y} axis that results from
\begin{widetext}
\begin{equation}
{\cal S}_{k_1\,k_2}(x|{\hat{\mathbf p}}(x)\,;\,x'|{\hat{\mathbf p}}(x')) =
\int d y' M_{k_1\,k_2}(x,y,{\hat{\mathbf p}(x)}; x',y',{\hat{\mathbf p}(x')})
\end{equation}
\end{widetext}
with \(M_{k_1\,k_2}\) minus the relevant Mayer function \cite{mayer}, 
which is equal to 
\(0\),  
if the two circular arcs of type, respectively, \(k_1\) and \(k_2\), and with position and orientation, respectively, 
\(x,y,{\hat{\mathbf p}(x)}\)  and \(x',y',{\hat{\mathbf p}(x')}\)  do not intersect, 
or \(1\), 
if they do (Fig. \ref{figura4}).

The calculation 
of \({\cal S}_{k_1\,k_2}(x|{\hat{\mathbf p}}(x)\,;\,x'|{\hat{\mathbf p}}(x')) \) 
is feasible but is not necessary.
It is sufficient to determine 
for which values of the variables \(x\) and \(x'\) 
the value of the function \({\cal S}_{k_1\,k_2}\) is equal to \(0\) and 
for which values of the variables \(x\) and \(x'\)  
the value of the function \({\cal S}_{k_1\,k_2}\) is larger  than \(0\).
Since hard (minor) circular arcs can intersect only if their parent circumferences intersect and
the present hard minor circular arcs 
translo-rotate on their parent semicircumferences and 
translate along the \textsl{y} axis,
one can realize that:
\begin{widetext}
\begin{equation}
{\cal S}_{k_1\,k_2}(x|{\hat{\mathbf p}}(x)\,;\,x'|{\hat{\mathbf p}}(x'))
\begin{cases}
= 0 \;,\; {\rm if} \; 0 \leq \left | x \right | \leq R \cos\left(\dfrac{\uptheta_{k_1}}{2}\right)  \; {\rm and} \; 
0 \leq \left | x' \right | \leq R \cos\left(\dfrac{\uptheta_{k_2}}{2}\right)  \\[10pt]
\geq 0 \;,\; {\rm else}
\end{cases}
\label{equformaS}
\end{equation}
\end{widetext}
(Fig. \ref{figura4}).

In the lowest extreme, mathematical, case \(\uprho = 0\),
irrespective of the form of \( {\cal S}_{k_1\,k_2}(x|{\hat{\mathbf p}}(x)\,;\,x'|{\hat{\mathbf p}}(x'))\),
the (free-energy) functional of Eq. \ref{equf} is minimized by 
the uniform functions 
\[G_k(x)=1\,,\] 
\(\forall k,\, k=1,\dots,{\rm n}\), and \(\forall \left\{\mathtt{x}_k\right\}_{\rm n} \in \mathtt{X}\).

In the highest extreme, physical, case \(\uprho \rightarrow \infty\),
the form of  
\( {\cal S}_{k_1\,k_2}
(x|{\hat{\mathbf p}}(x)\,;
\,x'|{\hat{\mathbf p}}(x'))\) 
of Eq. \ref{equformaS} implies 
that the free-energy functional of Eq. \ref{equf} is minimized by 
the non-uniform functions
\[
G_k(x) = 
\begin{cases}
\dfrac{1}{\cos\left(\dfrac{\uptheta_k}{2} \right)} \,\;,\; {\rm if} \; 0 \leq \left | x \right | \leq R \cos\left(\dfrac{\uptheta_k}{2}\right) \\
\\
0  \;,\; {\rm else}
\end{cases}
\,,
\] 
\(\forall k,\, k=1,\dots,{\rm n}\), and \(\forall \left\{\mathtt{x}_k\right\}_{\rm n} \in \mathtt{X}\).

The basic question to pose now becomes:
\(\exists \left\{\mathtt{x}_k\right\}_{\rm n} \in \mathtt{X}\) such that
are there strictly positive values of \(\uprho\), \(\uprho > 0\), 
for which
the uniform functions \(G_k(x)=1\), \(\forall k, \, k=1,\dots,{\rm n}\), 
can minimize  the free-energy functional of Eq. \ref{equf}?
To appreciate that the response is in the negative,
it is convenient to note that,
for any \(\left\{\mathtt{x}_k\right\}_{\rm n} \in \mathtt{X}\) and for any value of \(\uprho\), 
the functions \(G_k(x)\) that minimize the free-energy functional of Eq. \ref{equf}
are a solution 
of the correspondent set of integral equations:
\begin{widetext}
\begin{eqnarray}
\label{equint}
\log K_1 G_{k_1}(x) = - 
\uprho \sum_{k_2=1}^{\rm n} \mathtt{x}_{k_2} 
\int d x' \, G_{k_2}(x') \, {\cal S}_{k_1\,k_2}(x|{\hat{\mathbf p}}(x)\,;\,x'|{\hat{\mathbf p}}(x')), \\[-20pt]
\nonumber
\end{eqnarray}
\end{widetext}
\(\forall k_1,\, k_1=1,\dots,{\rm n}\), and 
with \(K_1\) the constant that ensures the normalization of \(G_{k_1}(x)\) via Eq. \ref{equnorma}. 

The basic question to pose now becomes:
except in the lowest extreme, mathematical, case \(\uprho = 0\),
\(\exists \left\{\mathtt{x}_k\right\}_{\rm n} \in \mathtt{X}\) such that
can the uniform functions \(G_k(x) = 1\), \(\forall k,\, k=1,\dots,{\rm n}\), be a solution 
of the set of integral equations of Eq. \ref{equint} 
for \(\uprho > 0\)?
To appreciate that the response is in the negative,
one can realize that, 
for \(G_k(x)=1\), \(\forall k, \, k=1,\dots,{\rm n}\), to be a solution,
the integrals in the right-hand side of Eq. \ref{equint} must be independent of \(x\), i.e.,
the functions \( {\cal S}_{k_1\,k_2} (x|{\hat{\mathbf p}}(x)\,;\,x'|{\hat{\mathbf p}}(x')) \) 
must depend only on the absolute difference \(\Updelta x = | x'-x |\).
However, because of the (strong) two--dimensional coupling between positions and orientations,
this is not the case:
\({\cal S}_{k_1\,k_2}(x|{\hat{\mathbf p}}(x)\,;\,x'|{\hat{\mathbf p}}(x'))\) 
depends on both \(x\) and \(x'\),
as one can appreciate from Eq. \ref{equformaS}.

Therefore, 
\(\forall \left\{\mathtt{x}_k\right\}_{\rm n} \in \mathtt{X}\), 
\(G_k(x) = 1\), \(\forall k, \, k=1,\dots,{\rm n}\), cannot be a solution 
of the set of integral equations of Eq. \ref{equint} for \(\uprho > 0\):
for any strictly positive value of \(\uprho\),
the artificial system, 
so designed as to be the best promoter of a ``splay(-bend)'' nematic phase,
has always to ``stratify'' to preserve perfect local polar order. 

There are a few objections that can be raised at this point.

\begin{figure}[h!]
\centering
\includegraphics[scale=0.65]{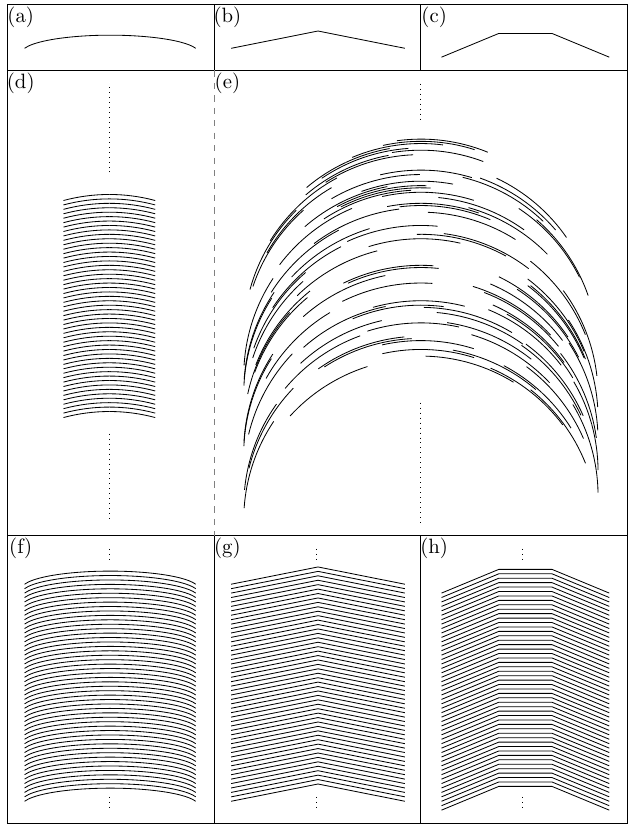}
\caption{Examples of other hard arcuate or bent infinitesimally--thin particles:
(a) elliptical arcs, (b) V-shaped needles, (c) bow-shaped needles.
Hard minor circular arcs can infinitely densely pack by both 
(d) stacking on top of one another and 
(e) disposing themselves on the parent semicircumferences. 
Instead,
(f) hard infinitesimally--thin elliptical arcs, (g) hard V-shaped needles and (h) hard bow-shaped needles (h)
presumably can infinitely densely pack only by stacking on top of one another.}
\label{figura5}
\end{figure}

The first objection concerns  
the hard infinitesimally--thin minor \textit{circular} arcs
being the most appropriate elementary model.
One could have considered other hard infinitesimally--thin particles,
such as hard infinitesimally--thin elliptical arcs [Fig. \ref{figura5}(a)] 
or hard V-shaped needles [Fig. \ref{figura5}(b)] 
or hard bow-shaped needles [Fig. \ref{figura5}(c)],
that would have been as elementary models.
Systems of hard V-shaped needles and 
systems of hard bow-shaped needles 
were actually investigated 
in the past by 
(Onsager) second-virial density functional theory and Monte Carlo numerical simulation
\cite{quintanaeuro,quintanastat,tavarone,polacchi1,polacchi2,polacchi3}.
In all of these investigations, 
the authors claimed to have found a ``modulated'' nematic phase,
which was termed ``bend'' \cite{quintanaeuro,quintanastat} 
or more generically ``modulated'' \cite{tavarone} 
or indeed ``splay-bend'' \cite{polacchi1,polacchi2},
except in the most recent investigation,
whose authors rectify their own past results \cite{polacchi3}.
The advantage to using minor circular arcs over 
elliptic arcs or V-shaped needles or bow-shaped needles 
resides in 
that only minor circular arcs, 
having being severed from a (semi)circumference and 
therefore having a constant curvature at any point of their curvilinear tract, 
allow that smooth translo-rotational motion 
that consists in adapting to their parent (semi)circumference [Fig. \ref{figura2}(a)].
Hard elliptical arcs cannot adapt to their parent ellipse, 
while the cuspid of the V-shaped needles obstruct such a smooth translo-rotational motion
and the cuspids of the bow-shaped needles, 
even though they are better proxys for minor circular arcs,
also obstruct such a smooth translo-rotational motion.
For exactly the same reason,
while hard infinitesimally--thin minor circular arcs 
can infinitely densely pack
by both stacking on top of one another
[Fig. \ref{figura5}(d)] and
disposing themselves on the parent semicircumferences
[Fig. \ref{figura5}(e)], 
hard infinitesimally--thin elliptic arcs, 
V-shaped needles and bow-shaped needles
presumably can infinitely densely pack 
only 
by stacking on top of one another 
[Fig. \ref{figura5}(f,g,h)].
This means that
(non-)polar layered  
(liquid-)crystalline phases 
will be the stablest phases for
uni-component systems of hard infinitesimally--thin elliptical arcs
and of V-shaped needles and of bow-shaped needles 
at sufficiently high density and 
therefore the region of stability of the ``modulated'' nematic phase,
if it existed in those systems of hard particles, 
would anyway be eroded from above in density 
by those positionally ordered phases,
while multi-component systems of hard infinitesimally--thin elliptical arcs
and of V-shaped needles and of bow-shaped needles can suffer arrest and/or fractionation
at sufficiently high density.

The second objection concerns the \textit{infinitesimal} thinness.
One could have considered hard finitely--thin minor circular arcs but 
so doing would have caused, in a uni-component system, the formation
of a positionally ordered (liquid-)crystalline phase 
at sufficiently high density 
that would have eroded 
the region of stability of a (``modulated'') nematic phase from above in density,
while, in a multi-component system, arrest and/or fractionation 
at sufficiently high density.
For exactly the same reason,
in the constitution of the artificial system,
one has considered hard arcuate or bent particles,
such as hard minor circular arcs,
and excluded hard cuneate particles 
since hard cuneate particles 
cannot but be finitely thin.
   
The third objection concerns the \textit{perfect} local polar order. 
Certainly this is a very artificial approximation but 
its abandonment would admittedly oblige the restoration of
the isotropic phase and the non-``modulated'', ordinary nematic phase, 
with their action of erosion of the region of stability 
of any liquid-crystalline phase, the former, 
and of a ``modulated'' nematic phase, the latter, 
from below in density.

Therefore,
even in the situation that should be the most favourable to its formation,
a ``splay(-bend)'', truly nematic, phase does not form.

It seems that ``modulation'', polarity and uniformity do not easily co-habit.
If one wants ``modulation'' and perfect local polarity then uniformity must be abandoned.
If one wants uniformity then 
there is always the possibility to abandon ``modulation''
and obtain the old non-``modulated'', ordinary, nematic phase,
in which also polarity has been abandoned,
or the new non-``modulated'' polar nematic phase \cite{nempola1,nempola2},
whose existence, however, is yet to be unquestionably confirmed.
If one does want both ``modulation'' and uniformity then
perfect local polarity, the would-be promoter of ``modulation'', 
has to be, rather paradoxically, abandoned.
However, one should accept the consequences of this abandonment.
It is untenable to adjust only polar order but constrain ``modulation'' and uniformity:
in such a delicate and subtle situation,
it is rigorous to allow any possible type of order to exist and 
let the system free to find the stablest phase at a certain value of density,
be it ``modulated'' or not, be it uniform or not.
Under these free and realistic conditions,
it is probable that the system will combine both manners
to lower its free energy:
introduction of non-uniformity and reduction of orientational (polar, nematic) order:
in fact, this is what has been observed in Monte Carlo numerical simulations 
and more complete
second-virial density functional theory calculations
on systems of hard minor circular arcs \cite{hardarcs}.
 
The present considerations and results join 
the previous less recent results on systems of hard spherical caps \cite{hardsphericalcaps} and
the previous more recent results on systems of  hard minor circular arcs \cite{hardarcs}  
to indicate the inexistence of a ``splay(-bend)'' nematic phase:
rather,
what one is dealing with in systems of (hard) curved particles
is a phenomenon of (entropic) clusterization 
\citesx{hardsphericalcaps,hardarcs}\citedx[and also][]{hardarcsdp,hardcapsids}.
Consequently,
the present considerations and results join
the previous less recent results on systems of hard spherical caps \citevv{hardsphericalcaps} and
the previous more recent results on systems of hard minor circular arcs \cite{hardarcs}
to suggest a revision 
of those experimental investigations
that claim to have observed a ``splay(-bend)'' nematic phase 
\citesx{goodby}\citedx[and also][]{sciencepaesibassi}.



\end{document}